# Skew Adjustment Factors for Fragilities of California Box-Girder Bridges Subjected to Near-Fault and Far-Field Ground Motions

Sujith Mangalathu, Ph.D.[1], Jong-Su Jeon, Ph.D.[2]*, and Jiqing Jiang, Ph.D.[3]

[1]Post-Doctoral Fellow, Department of Civil and Environmental Engineering, University of California, Los Angeles, CA 90095, USA. Email: sujithmangalath@ucla.edu

[2]Assistant Professor, Department of Civil Engineering, Andong National University, Andong, Gyeongsangbuk-do 36729, Republic of Korea. Email: jsjeon@anu.ac.kr

[3]Associate Professor, Department of Civil Engineering, Zhejiang University City College, Hangzhou 310015, China. Email: jiangjq@zucc.edu.cn

* Corresponding author

**Abstract**

Past reconnaissance studies revealed that bridges close to active faults are more susceptible to damage and more than 60% of the bridges in California are skewed. To assess the combined effect of near-fault ground motions and skewness, this paper evaluates the seismic vulnerability of skewed concrete box-girder bridges in California subjected to near-fault and far-field ground motions. The relative risk of skewness and fault-location on the bridges is evaluated by developing fragility curves of bridge components and system accounting for the material, geometric, and structural uncertainties. It is noted that the skewness and bridge site close to active faults make bridges more vulnerable, and the existing modification factor in HAZUS cannot capture the variation in the median value of the fragilities appropriately. A new set of fragility adjustment factors for skewness coupled with the effect of fault location is suggested in this paper.

**Introduction**

As fragility curves are implemented in earthquake situational awareness application such as ShakeCast (Wald et al. 2008), it is critical to develop reliable fragility curves of structures. The implementation of the most reliable fragility curves helps assess the impact of earthquakes on the critical lifeline facilities (e.g. bridges), and assist the emergency responders to have informed decision on the recovery and operational strategies of the infrastructural systems after an earthquake (Mangalathu 2017). Fragility curves are conditional probability statements that give the likelihood of damage in a structure as a function of the ground motion intensity measure (IM). These curves can account for the uncertainties in ground motions, geometric, structural and material properties of the structure.

Following the disastrous earthquakes such as the 1971 San Fernando and 1989 Loma Prieta earthquakes, extensive studies have been performed to derive fragility curve of bridges in California. Although initial studies on the generation of fragility curves for bridges in California were focused on empirical fragility curves (Shinozuka et al. 2000), fragility curves have been generated using extensive numerical analyses over the last decades (Basöz and Mander 1999; Brandenberg et al. 2011; Gardoni et al. 2002; HAZUS-MH 2003; Huo and Zhang 2013; Mackie and Stojadinovic 2001; Mangalathu 2017; Mangalathu and Jeon 2018; Mangalathu et al. 2018a; Ramanathan 2012; Zhong et al. 2009). Although HAZUS is the only document, which suggests fragility relationships for all the bridge classes in California, the HAZUS methodology and fragility relationships are criticized by recent researches (Mangalathu et al. 2017b; Porter 2010; Ramanathan 2012). The downside of the above cited studies is that the studies did not address the effect of near-fault (NF) and far-field (FF) ground motions separately. NF ground motions may have more distinct characteristics than FF ground motions (e.g., directivity, spectral non-stationarity, intensity, duration, frequency content characteristics, and directionality of components) and have significant influence on the seismic response of bridges as realized in the recent bridge reconnaissance. (Chang et al. 2000; Loh et al. 2002), analytical (Billah et al. 2012; Dimitrakopoulos 2011; Jalali et al. 2012; Liao et al. 2004; Ozbulut and Hurlebaus 2012; Shen et al. 2004; Shi and Dimitrakopoulos 2017), and experimental studies (Choi et al. 2010; Phan et al. 2007; Saaidi et al. 2012).

Liao et al. (2004) compared the dynamic response of box-girder bridges in Taiwan, and noted that the ratio of peak ground velocity to acceleration is the key factor that governs the response of bridges under NF ground motions. Shen et al. (2004) investigated the performance of isolated bridges in Taiwan under NF motions. Their dynamic analysis revealed that the seismic response of bridges is amplified when the pulse period is close to the effective period of the isolation system. Park et al. (2004) investigated the seismic performance of the Bolu Viaduct in Turkey consisting of yielding-steel energy dissipation units and sliding pot bearings. Jónsson et al. (2010) pointed out from an analytical study that the damage on a base-isolated bridge in Iceland was due to the neat-fault effect. Dimitrakopoulos (2011) noted that under the influence of NF motions, the tendency of skew bridge to rotate after pounding is not a factor of the skew angle alone, but the whole geometry in plan and the friction. Ozbulut and Hurlebaus (2012) examined the effect of superelastic-friction base isolator to reduce deck drift. These authors concluded that the re-centering capability due to superelastic-friction base isolator reduced the residual defamation of bridges. All these studies (Jónsson et al. 2010; Liao et al. 2004; Ozbulut and Hurlebaus 2012; Park et al. 2004; Shen et al. 2004) were focused on the effect of isolation systems of bridges under NF ground motions, and have not investigated the effect of NF and FF ground motions on the bridge fragilities. Due to the discrepancies in design details (Mangalathu 2017), these studies are not helpful in assessing the seismic performance of bridges in California.

The investigation of NF effects on bridge fragilities requires special attention as 73% of the bridges in California are close to known active faults (Choi et al. 2010).

Some analytical and experimental studies have been conducted to examine the impact of NF and FF ground motions on bridges in California. Billah et al. (2012) investigated the effect of various retrofitting strategies on bridges in North America under NF and FF ground motions, and concluded that the bridges are more vulnerable to NF ground motions. Kaviani et al. (2012) examined the seismic demand analysis for specific two-span and three-span seat abutment bridges in California. Their work revealed that the high velocity pulses increase the seismic demand of the bridges and that the skew angle increases the collapse potential of the bridges. Zakeri et al. (2014a) pointed out from the fragility analysis of two-span box-girder bridges that the existing skew modification factors presented in HAZUS can reflect the impact of skew on the fragility satisfactorily. From the analytical evaluation of various retrofitting strategies for two-span bridges in California constructed before 1971, Zakeri et al. (2014b) noted that the efficiency of the retrofitting strategy varies depending on the skew angle. Omrani et al. (2017) noted from the study of a skewed two-span two-column bent bridge with seat abutments that the adopted model for the abutments have a significant influence on the system and component fragilities. However, all of the above studies did not fully address the fragility difference of skewed bridge classes subjected to FF and NF ground motions.

The shake table test on two bridge columns by Phan et al. (2007) noted that NF ground motions increase the residual displacement of bridge columns. Bridges with a residual drift of more than 1.75% were demolished following the 1995 Kobe earthquake although they did not collapse during the earthquake (Ardakani and Saiidi 2013; Cheng et al. 2016; Kawashima et al. 1998). However, most of the fragility studies of bridges in California (e.g., Mangalathu et al.

2016; Mangalathu 2017; Ramanathan 2012) are limited to the peak response of bridge components than residual deformation. It is thus noted that further studies are needed to account for the effect of NF ground motions and residual displacement on the seismic vulnerability of bridges in California.

Another contribution of this research is to examine the impact of NF ground motions on the fragility curves for skewed bridges in California. Extensive studies have been conducted to investigate the effect of skewness (Amjadian et al. 2016; Dimitrakopoulos 2011; Huo and Zhang 2013; Maleki 2002; Mangalathu et al. 2018; Meng et al. 2004; Ramanathan et al. 2015) on the bridge responses. However, studies on examining the effect of NF ground motions on structural responses are yet scarce. Shamsabadi et al. (2004) noted that the strong velocity pulse has a significant effect on the skewed bridges in California. These authors pointed out the need for an extensive study on the response of skewed bridges under NF ground motions. Such a research is critical as skewed bridges occupy more than 60% of the California bridge inventory (Mangalathu et al. 2018).

Based on the knowledge gap noted in literature review, the objective of this paper is multifold: (1) to examine the effect of NF and FF ground motions on the seismic vulnerability of skewed multi-span concrete box-girder bridges in California reflecting the material, geometric, and structural uncertainties (bridge-class fragility characteristics), (2) to propose component and system adjustment factors for bridge-fragilities accounting for the skewness, NF and FF effects, (3) to suggest demolition fragilities for skewed multi-span concrete box-girder bridges in California, and (4) compare the proposed fragility relationships with HAZUS fragility relationships. To achieve these sub-tasks, this research selects three-span single-column bent bridge classes with diaphragm and seat abutments in California as these bridges occupy more than 25% of the California bridge inventory (Mangalathu 2017; Ramanathan 2012). The generation of fragility curves for the entire bridge inventory in California accounting for the earthquake type and skew angle is beyond the scope of the current study. However, the insights from the current study are useful in future research. Numerical models are generated in OpenSees (McKenna 2011) including the material, geometric and structural uncertainties. For the selected NF and FF ground motion sets, nonlinear time history analyses (NLTHAs) are performed to generate probabilistic seismic demand models (PSDMs) and fragility curves of skewed bridges. Fragility curves using NF and FF ground motions are compared and are used to propose the skew adjustment factors for NF and FF effects.

**A Suite of Near-Fault and Far-Field Ground Motions**

NF ground motions may have distinct characteristics such as the rupture directivity effect in the fault-normal direction and a permanent displacement in the fault-parallel direction (Dabaghi 2014). Directivity (dependence on the rupture direction) is one of the primary factors affecting motion at a near-fault site. Forward directivity happens when the fault rupture propagates towards the site and the ground motions exhibit a large velocity pulse. The forward directivity is generally characterized by the presence of a two-sided, large-amplitude velocity and long-period pulse in the fault-normal direction. Backward directivity occurs when the fault rupture propagates away from the site, and is characterized by low intensity and long duration pulse. Recently, extensive efforts have been carried out to account for the directivity effect in the ground motion models (Dabaghi 2014; Shahi and Baker 2014; Spudich et al. 2014). This research uses the recorded 120 pairs of NF ground motions summarized by Dabaghi (2014) for the generation of fragility models. 120 pairs of the FF ground motions suggested by Baker et al.

(2011) for the PEER transportation program is employed as FF ground motions in this research. All ground motions are scaled by a factor of 1.5 and two to have sufficient response data with respect to various ranges of IMs (Ramanathan 2012). Thus, a total of 360 ground motions are used for this research for each suite of NF and FF ground motions. To be consistent with HAZUS fragility relationships, the spectral acceleration at a period of one second ($S_{a-1s}$ in g) is adopted as the IM in this research. Ramanathan (2012) indicated that $S_{a-1.0s}$ is the optimal intensity measure for California concrete box-girder bridge classes based on the combined effect of near-field and far-field ground motions. A further study is needed to evaluate the sufficiency of this IM for near-fault ground motions alone. The effect of vertical acceleration is not considered in the current study.

**Numerical Modeling and Fragility Methodology**

This research selects three-span concrete box-girder bridges with single-column bents constructed after 1970s. The abutments at the ends can be of diaphragm or seat. Diaphragm abutments are cast monolithic with the superstructure while seat abutments provide a bearing support to the superstructure. The finite element platform OpenSees (McKenna 2011) is used to model the bridge configurations. Rayleigh damping is adopted in dynamic analyses for the first and second vibration modes. The bridge superstructure is modeled as elastic using the elastic beam-column elements, as shown in Fig. 1. Transverse deck elements are assumed to be rigid and are connected to the columns using rigid elements to ensure the moment and force transfer between adjacent components. Fiber-type displacement-based beam-column elements are used to simulate the nonlinear behavior of the columns. Translational and rotational springs are added at the column base to simulate the behavior of the footing, and are modeled using linear elastic elements. For seat abutments, the expansion joint between the deck and an abutment consists of various components such as elastomeric bearings (longitudinal and transverse), shear key (transverse), and pounding between the deck and abutment (perpendicular to the backwall). The response of bearing elements is simulated using a bilinear model, and the pounding behavior is simulated using the model suggested by Muthukumar and DesRoches (2006). The friction model suggested by Muthukumar and DesRoches (2006) is of bi-linear nature that can capture the impact and energy dissipation (Hertz contact model). The effect of frictional contact force (Saiidi et al. 2012) is not considered in the study and interested readers are directed to Shi and Dimitrakopoulos (2017) for a more advanced pounding model. Based on the experimental results conducted by Silva et al. (2009), the shear key is modeled as trilinear with gap (Fig. 1).

The soil and pile springs are rotated with respect to the abutment skew. To account for this abutment skew, the soil model developed by Shamsabadi et al. (2010) is modified under the assumption that the direction of the passive pressure is perpendicular to the backwall plane. Following the work of Kaviani et al. (2012), the variation coefficient of stiffness and strength for a specified skew angle is defined as $0.3 \cdot \tan(\alpha)/\tan(60°)$. The upper limit of this coefficient is 0.3 at a skew angle of 60°. It is also assumed that active resistance of the abutment is contributed by the piles alone. More detailed descriptions on the numerical modeling of bridge components are provided in the references (Jeon et al. 2016; Mangalathu 2017; Mangalathu et al. 2017a).

Different sources of uncertainties, such as geometric, material, and system, are included in this research. Table 1 presents the mean value ($\mu$), standard deviation ($\sigma$), and the associated probability distribution of various input variables used in this research. The values are determined based on an extensive plan review of bridges (more than 1,000), as reported in Mangalathu (2017). This research accounts for the statistical dependence of superstructure and span length, span length and column diameter in generating the bridge samples (Mangalathu 2017). To reflect all the possible bridge configurations in California, other input parameters are randomly sampled. However, special quality assurance is carried out to ensure the realistic nature of bridge samples. For each parameter, the samples greater/smaller than 1.96 times the standard deviation from its mean value (95 confidence interval) are regarded as outliers and are not considered in this research (truncation limits, Table 1). The sensitivity of the input parameters on bridge fragilities is reported in Mangalathu et al. (2018b).

Note that the geometric, material, and structural uncertainties are considered in the current study as the intention of the current study is to generate fragility curves based on bridge inventory for regional seismic risk assessment. The consideration of these uncertainties used for developing the fragility curve for the selected bridge classes is consistent with previous researches on fragility curves developed for regional risk assessment (Ramanathan 2012; Mangalathu 2017). Consistent with 320 ground motions, 320 statistically significant yet nominally identical bridge models are generated and are randomly paired with the ground motions. The bridge models are generated by sampling across the parameters using Latin Hypercube sampling technique (LHS), and LHS provides an effective scheme to cover the probability space of the random variables in comparison to pure random sampling using naïve Monte Carlo Simulation (McKay, 1979). Also, Mangalathu (2017) noted that 320 bridge samples can capture the uncertainties associated with the input variables. NLTHA is carried out on the ground motion-bridge pair to monitor the response of various components (defined as engineering demand parameters (EDPs)). Seven EDPs are used in this research: the maximum column drift ($\theta_c$ in %, COL), residual drift of column ($\theta_R$ in %, COLR), maximum passive abutment displacement ($\delta_p$ in mm, ABP), maximum active abutment displacement ($\delta_a$ in mm,

ABA), maximum tangential abutment displacement ($\delta_t$ in mm, ABT), maximum deck unseating displacement ($\delta_u$ in mm, UST), and maximum bearing displacement ($\delta_b$ in mm, BRG). Demand models the selected bridge class are generated based on NTLHA results for 320 bridge-ground motion pairs. Linear regression analysis is conducted on the demand (D) and intensity measure (IM) in a logarithmic space to generate the PSDM of bridge components (Cornell et al. 2002):

$$\ln(S_d) = \ln(a) + b\ln(IM) \qquad (1)$$

where $a$ and $b$ are the regression coefficients, $S_d$ is the median estimate of the demand in terms of $IM$. The coefficients $a$ and $b$ are obtained through a linear regression analysis on $D$ and $IM$ pairs in the logarithmic space. Dispersion, $\beta_{d|IM}$, is evaluated based on statistical analysis of $D$ and $IM$ pairs:

$$\beta_{d|IM} = \sqrt{\frac{1}{N-2}\sum_{i=1}^{N}[\ln(d_i) - \ln(S_d)]^2} \qquad (2)$$

where $d_i$ is the demand for the $i$th ground motion.

Assuming that both demands and capacities follow a lognormal distribution, the fragility function for a bridge component is defined as a lognormal cumulative distribution function:

$$P[D > C | IM] = \Phi\left[\frac{\ln(S_d / S_c)}{\sqrt{\beta_{d|IM}^2 + \beta_c^2}}\right] \qquad (3)$$

where $S_d$ and $\beta_{d|IM}$ are the median and dispersion, respectively, of the demand conditioned on $IM$. $\Phi[\bullet]$ is the standard normal cumulative distribution function. $S_c$ and $\beta_c$ are the median and dispersion, respectively, of the capacity or limit states. A set of component fragility curves computed in Eq. (1) has to be integrated to a system fragility (or bridge fragility), which is facilitated through the development of joint probabilistic seismic demand models (JPSDMs) (Nielson 2005; Mangalathu 2017). The JPSDM recognizes the correlation between various components. If the vector demands, $X_i$, placed on the $n$ components of the system are expressed as $\underline{X} = (X_1, X_2, ..., X_n)$, then the vector, $\underline{Y} = \ln(\underline{X})$ represents the vector of component demands in the log-transformed space. The JPSDM is formulated in this space by assembling the vector of means $\mu_{\underline{Y}}$ and the covariance matrix, $\sigma_{\underline{Y}}$. Monte Carlo simulation is then carried to compare the demand and capacity realizations. Following previous studies (Nielson 2005; Mangalathu 2017), $10^6$ demand and capacity samples are used to estimate the probability that the demand exceeds the associated capacity value for each *IM*. This procedure is repeated for the increasing value of the *IM*, and regression analysis is used to estimate the lognormal parameters, median and dispersion, which characterize the bridge fragility. Additionally, the series system assumption is considered in the current study due to the fact that the any component level damage induces the similar system level damage. Interested readers are directed to the references (Mangalathu 2017; Nielson 2005) for a more detailed description of the fragility methodology. As presented in Table 2, the limit states of all the bridge components except for the residual column deformation follow the lognormal distribution, following the work of Dutta (1999) and Mangalathu (2017). Note that the limit states in Table 2 are aligned with Caltrans design and operational experience that facilitates the evaluation of repair-related decision variables, repair cost, repair time and traffic implications (Mangalathu, 2017). Since it is very difficult to quantify the residual deformation of the columns, this research regards the limit state of the residual column deformation as deterministic for demolition. Thus, to develop the demolition fragility curve, this research adopts logistic regression for the generation of demolition fragility curves of bridges, and is explained in the next section.

**Comparison of Near-Fault and Far-Field Fragilities**

To examine the effect of skew angle ($\alpha$) on the seismic vulnerability of the selected bridge classes under NF and FF ground motions, their component and system fragility curves are generated with different levels of skew angle. Here, four levels of skew angle ranging from normal to high skew angle are selected: $\alpha = 0°, 15°, 30°,$ and $45°$.

*Comparison Based on Maximum Response*

Fragility curves for diaphragm and seat abutment bridge classes are generated by convolving PSDM based on the maximum response of various bridge components and the limit states presented in Table 2. Tables 3 and 4 show the median value ($\lambda$) and dispersion ($\beta$) of fragility curves for the selected bridge classes under NF and FF ground motions. As noted from Tables 3 and 34, the column is not always the most vulnerable component that governs the system fragility. The most vulnerable component can be the bearing, abutment, or column depending on the limit state and the skew angle. Note that the relative vulnerability of the bridges is evaluated in this research through the change in the median value of the fragility curves. Following inferences are obtained from the comparison of bridge fragility values using NF and FF ground motions with various skew angles:

- Bridges subjected to NF ground motions are more vulnerable than FF ground motions. In the case of diaphragm abutment bridge class with zero skew, the change in median values between the NF and FF fragilities are 12%, 49%, and 62%, respectively, for $LS_2$, $LS_3$ and $LS_4$. For seat abutment bridge class, bridges subjected to NF ground motions are 6%, 7%, 45%, and 59% more vulnerable than under FF ground motions for $LS_1$ through $LS_4$. The increased vulnerability of bridges subjected to NF ground motions is attributed to the increased demand of the bridge components due to the NF ground motions. It is also noted that difference in the median value of fragilities between the NF and FF fragilities increases with the increase in limit states. Although not shown here, the conclusion holds true for bridges with different skew angles. It highlights the need for updated design requirements for bridges located in near-fault areas suggested by Ardakani and Saiidi (2013). It also underscores the need to suggest NF and FF bridge fragilities in further version of HAZUS.

- For diaphragm abutment bridge class, the transverse abutment displacement governs the system fragilities at lower limit states ($LS_1$ and $LS_2$), while for seat abutment bridge class, the bearing deformation controls the system fragilities at the same limit states. This difference is associated with the load transfer mechanism due to the existence of the bearings. For both bridge classes, the column drift, not the deck unseating, governs the system fragility at higher limit states ($LS_3$ and $LS_4$). As shown in Table 2, the column and unseating are only the primary components affecting extensive damage and bridge collapse. However, most of the recorded unseating deformations are smaller than the seat width (unseating capacity) in Table 2, and thus the unseating does not contribute significantly on the bridge system. The conclusion is valid for NF and FF bridges, and is consistent with the previous studies on the fragility analysis of bridges in California (Mangalathu 2017; Mangalathu and Jeon 2018).

- The bridge becomes more vulnerable with the increase in skew angle. In the case of seat abutment bridges subjected to FF ground motions, the 45º skewed bridges are 7%, 1%, 27% and 29% more vulnerable than the non-skewed bridges, respectively, for $LS_1$ through $LS_4$. For seat abutment bridges under NF ground motions, the change in median value between the 45º skewed and non-skewed bridges are 8%, 3%. 20% and 20%, respectively, for $LS_1$ through $LS_4$. It is due to the fact that bridges tend to rotate during an earthquake with the increase in the skew angle. This in-plane rotation induces the high seismic demand of the

- bridge components such as bearing, deck displacement, transverse abutment displacement, and column drift, which makes the bridge more vulnerable.
- The rate of variation of the median value of fragilities with skew angle is different for different components. For example, in the case of diaphragm abutment bridge class under FF motions, the change in skew angle from 0º to 45º decreases the median value of the maximum column drift for LS$_2$ by 18%, while for the active abutment action, the decrease in median value is 80%. Such an observation points the necessity of different modification factors for different components as a function of skewness.
- The dispersion ($\zeta$) of the fragility relationships varies depending on the bridge components under consideration and the constant dispersion of 0.35 suggested by HAZUS needs further revision.

*Comparison Based on Residual Drift*

Following the 1995 Kobe earthquake, more than 100 bridges with residual drift of more than 1.75% were demolished although the bridges were not collapsed (Kawashima et al. 1998). However, past fragility studies for bridges in California (HAZUS-MH 2003; Mangalathu et al. 2016; Mangalathu 2017; Ramanathan 2012) were focused mainly on the maximum column demand than the residual column deformation. The residual deformation makes the columns difficult to repair and thus the bridge is usually demolished (Kawashima et al. 1998). A residual drift of 1.75% is used in this research to generate the demolition fragility curves and are generated based on the logistic regression of the IM and the failure-survival vector of residual drift; residual drift exceeding 1.75% is marked as failure and less than 1.75% is marked as survival. Fig. 2 shows the demolition fragilities of diaphragm and seat abutment bridge classes with various degrees of skew angle under NF and FF ground motions. Following inferences can be drawn from the comparison of fragilities presented in Fig. 2.

- In the case of diaphragm and seat abutment bridge classes subjected to FF ground motions, the skew angle increases the demolition vulnerability of bridges; the non-skewed bridge is less vulnerable than the skewed bridges. However, this conclusion is not true for the bridges subjected to NF ground motions. The presence of skew angle generally increases the demolition vulnerability for seat abutment bridge class but no specific pattern between skew angle change and vulnerability. Logistic regression depends on the number of failed simulations and failure-induced intensity measure values. For FF ground motions, the number of failed simulations for skewed bridges is slightly greater than that for non-skewed bridges (all skewed bridges fail at the same earthquakes) and the failure occurs at the same intensity measure. On the other hand, for NF ground motions, the number of failed simulations is different for four skewed cases and earthquakes producing the demolition state are different with respect to skew angle.
- Bridges subjected to NF ground motions are more vulnerable to demolition than those under FF ground motions. The conclusion holds true for both bridge classes.

**Skew Adjustment Factors for Near-Fault and Far-Field Fragilities**

As this research highlights the need for different adjustment factors for different bridge components, adjustment factors for the median value of the bridge fragilities as a function of skew angle is developed using the fragility characteristics in Tables 3 and 4. The adjustment factors are suggested based on the least squares fitting technique (linear or quadratic function for simplicity) on the change in regression coefficients with respect to the change in skew angle. Fig.

3 shows the regression model of the maximum column drift (COL) and active abutment displacement (ABA) for seat abutment bridge class at $LS_1$ under FF ground motions. It is noted that the suggested adjustment equations have good predict capabilities with a coefficient of determination ($R^2$) of 0.997 and 0.987, respectively, for the column drift and active abutment displacement. Tables 5 to 8 presents the adjustment factors for various bridge components as a function of skew angle. Note that the adjustment factor varies depending on the component, limit state, and bridge site under consideration. It is also noted that there is not much statistical variation on the dispersion between the fragilities for different skew angles and thus a constant value is suggested.

To evaluate the accuracy and efficiency of the proposed fragility modification factors, fragility curves based on the adjustment factors and are compared with the simulation-based (not adjusted) fragility curves. Fig. 4 shows the comparison of adjustment-based and simulation-based system fragilities at $LS_2$ and $LS_3$ for diaphragm and seat abutment bridge classes under NF ground motions. Note that the adjustment-based fragility curves are generated using the adjustment factors suggested in Tables 5 to 8, and the adjustment factors are derived based on results of NTHLAs. It is noted from the comparison of the fragility curves that there is no or little statistical variation between the two fragility curves, and the proposed modification factor can be used for the generation of fragility curves for different skew angles without extensive numerical simulations.

**Comparison of HAZUS and Proposed Fragility Relationships**

Fig. 5 shows the comparison of HAZUS fragilities with the proposed (adjustment-based) fragility relationships of the selected bridge classes under NF and FF ground motions. Note that HAZUS fragility relationships (1) assume that the bridge vulnerability is governed by the column response and (2) does not reflect the material, geometric, and structural uncertainties in the generation of fragility curves. Interested readers are directed to Mangalathu et al. (2017b) for a critical review of the HAZUS fragility relationships. As noted before, HAZUS suggests same fragility relationships for NF and FF ground motions as well as for diaphragm and seat abutment bridges. Following inference can be deduced from Fig. 5.

- HAZUS considerably underestimates the median value of the fragility, and thus the seismic risk of three-span bridge classes at lower limit states ($LS_1$ and $LS_2$). However, HAZUS significantly overestimates the seismic risk of three-span bridge classes at higher limit states ($LS_3$ and $LS_4$). This conclusion holds true for both seat and diaphragm abutment three-span bridges

- HAZUS can capture the general trend that the skew makes the bridges more vulnerable.

- The same fragility realtionships suggested by HAZUS for seat and dispahragm abutment bridges is not realistic, and for the selected bridge classes, seat abutment bridges are more vulnerable than diaphgram abutment bridges.

- For HAZUS fragility curves, as the skew angle increases, the median value of fragility curves decreases and the bridge vulnerability increases. However, for the proposed fragility curves, the skew angle increases the bridge vulnerability at higher limit states regardless of abutment type, which is associated with the increase of the column vulnerability (governing failure mode). On the other hand, the bridge vulnerability does not necessarily increases with the increase of skew angle at lower limit states. This is due to the fact that other components such as the transverse abutment action (for diaphragm abutments) and bearing (for seat abutments) govern the system vulnerability.

**Conclusions**

Past earthquakes have revealed that bridges close to fault are more susceptible to damage and more than 73% of the bridges in California are located close to known active faults. This paper compares the seismic vulnerability of skewed bridge classes in California subjected to near-fault (NF) and far-field (FF) ground motions. To achieve this aim, two types of bridge are included in this research: three-span single-column bent bridges with seat abutments and diaphragm abutments. The seismic vulnerability of the selected bridge classes is evaluated by developing fragility curves accounting for the material, geometric, structural, and ground motion uncertainties. Fragility curves are generated for various bridge components such as column, abutment actions in active, passive and transverse direction, unseating, and bearing and for the bridge system. This paper also evaluates the effect of skew on the bridges located close to active fault and far from active fault. The salient features noted from this research are:

- Bridges subjected to NF ground motions are more vulnerable than those under FF ground motions. The difference in the median value of fragilities increases with the increase in the limit states.

- The bridge becomes more vulnerable with the increase in skew angle, and the rate of variation of the median value of fragilities with skew angle is different for different components.

- For diaphragm and seat abutment bridge classes, the transverse abutment action governs the system fragilities at lower limit states (slight and moderate), while the maximum column drift governs the system fragility at higher limit states (extensive and complete).

- The skew angle increases the demolition vulnerability of the bridges when diaphragm and seat abutment bridge classes are subjected to FF ground motions.

This research also noted that the modification factor suggested by HAZUS cannot capture the variation in the median value of fragilities with skew angle appropriately. New adjustment factors are suggested in this paper as a function of skew angle for various bridge components for NF and FF ground motions. It is noted from the comparison of the fragility curves generated by the adjustment factor with the simulation-based (not adjusted) fragility curves that there is no or little variation between the fragility curves, and the proposed modification factor can be used to generate fragility curves for different skew angles without extensive computational efforts.

Compared to existing fragility relationships in HAZUS, the fragility curves presented in this paper are more reliable and can accurately represent the seismic vulnerability of the selected bridge configurations. The proposed fragility modification factors help the emergency responders to have a more reliable and informed post-earthquake recovery decision. The current study does not account for the effect of friction in the rotation of deck after pounding and further studies will be performed to examine the influence of this factor on bridge fragilities coupled with the effect of skew, ground motion type, and other bridge configurations.

**Acknowledgements**

This research was supported by Basic Research Program in Science and Engineering through the National Research Foundation of Korea funded by the Ministry of Education (NRF-2016R1D1A1B03933842).

**List of Tables**

**Table 1** Uncertainty Parameters of Bridges and Their Probability Distribution (Mangalathu 2017)

| Parameter | Type§ | Parameters | | Truncated limit | |
|---|---|---|---|---|---|
| | | Mean ($\mu$) | Standard deviation ($\sigma$) | Lower | Upper |
| **Superstructure (pre-stressed concrete)** | | | | | |
| Main-span length, $L_m$ (m) | N | 47.24 | 13.72 | 20.36 | 74.13 |
| Ratio of approach-span to main-span length, ($\eta = L_s/L_m$) | N | 0.75 | 0.2 | 0.4 | 1.0 |
| Width of the deck, $D_w$ (m) (three-cell deck) | N | 12.80 | 0.61 | 11.80 | 13.80 |
| **Interior bent** | | | | | |
| Concrete compressive strength, $f_c$ (MPa) | N | 31.37 | 3.86 | 23.80 | 38.94 |
| Rebar yield strength, $f_y$ (Mpa) | N | 475.7 | 37.9 | 401.4 | 550.1 |
| Column clear height, $H_c$ (m) | LN | 7.13 | 1.15 | 4.88 | 9.38 |
| Column diameter, $D_c$ (1.524 m vs. 1.676 m) | B | – | – | – | – |
| Column longitudinal reinforcement ratio, $\rho_l$ | U | 0.02 | 0.006 | 0.01 | 0.03 |
| Column transverse reinforcement ratio, $\rho_t$ | U | 0.009 | 0.003 | 0.004 | 0.013 |
| **Deep foundation (pile group)** | | | | | |
| Translational stiffness, $K_{ft}$ (kN/mm) | LN | 352.8 | 42.53 | 276.8 | 443.2 |
| Transverse rotational stiffness, $K_{fr}$ (GN-m/rad) | LN | 9.23 | 1.93 | 6.03 | 13.56 |
| Transverse/longitudinal rotational stiffness ratio, $K_r$ | LN | 1.53 | 0.32 | 1.0 | 2.25 |
| **Exterior bent** | | | | | |
| Diaphragm abutment backwall height, $H_a$ (m) | LN | 3.39 | 0.69 | 2.20 | 4.92 |
| Diaphragm pile stiffness, $K_p$ (kN/mm) | LN | 0.093 | 0.033 | 0.044 | 0.174 |
| Seat abutment backwall height, $H_a$ (m) | LN | 3.59 | 0.65 | 2.48 | 5.03 |
| Seat pile stiffness, $K_p$ (kN/mm) | LN | 0.124 | 0.045 | 0.059 | 0.232 |
| Backfill type, $BT$ (sand vs. clay) | B | – | – | – | – |
| **Bearing (elastomeric bearing)** | | | | | |
| Stiffness per deck width, $K_b$ (N/mm/mm) | LN | 1.4 | 0.779 | 0.448 | 3.439 |
| Coefficient of friction of bearing pad, $\mu_b$ | N | 0.3 | 0.1 | 0.1 | 0.5 |
| **Gap** | | | | | |
| Longitudinal (pounding), $\Delta_l$ (mm) | LN | 23.3 | 12.4 | 7.8 | 55.4 |
| Transverse (shear key), $\Delta_t$ (mm) | U | 19.1 | 11.0 | 0 | 38.1 |
| **Other parameters** | | | | | |
| Mass factor*, $m_f$ | U | 1.05 | 0.06 | 0.95 | 1.15 |
| Damping ratio, $\xi$ | N | 0.045 | 0.0125 | 0.02 | 0.07 |
| Acceleration for shear key capacity (g), $a_{sk}$ | LN | 1.0 | 0.2 | 0.8 | 1.2 |
| Earthquake direction (fault normal FN vs. parallel FP), $ED$ | B | – | – | – | – |

§ N = normal, LN = lognormal, U = uniform, and B = Bernoulli distribution.

* Mass factor presents the presence of parapets and barrier rails, variable deck slab thickness, electric poles, other equipment, etc.

**Table 2.** Limit State Models of Various Bridge Components

| Component | Median value, $S_c$ | | | | $\beta_c$ | Demolition |
|---|---|---|---|---|---|---|
| | Slight (LS$_1$) | Moderate (LS$_2$) | Extensive (LS$_3$) | Complete (LS$_4$) | | |
| Column drift (%) | | | | | | |
|   Maximum drift (COL) | 1.0 | 2.5 | 5.0 | 7.5 | 0.35 | |
|   Residual drift (COLR) | | | | | | 1.75 |
| Abutment deformation (mm) | | | | | | |
|   Passive action (ABP) | 76 | 254 | – | – | 0.35 | |
|   Active action (ABA) | 38 | 102 | – | – | 0.35 | |
|   Tangential action (ABT) | 25 | 102 | – | – | 0.35 | |
| Deck unseating (mm) (UST) | – | – | 254 | 381 | 0.35 | |
| Bearing displacement (mm) (BRG) | 25 | 102 | – | – | 0.35 | |

**Table 3.** Fragilities for Diaphragm Abutment Bridge Class for Different Skew Angles

| Skew angle | | Under FF ground motions | | | | | Under NF ground motions | | | | |
|---|---|---|---|---|---|---|---|---|---|---|---|
| | | $\lambda$ | | | | $\zeta^1$ | | $\lambda$ | | | $\zeta$ |
| | | LS$_1$ | LS$_2$ | LS$_3$ | LS$_4$ | | | LS$_1$ | LS$_2$ | LS$_3$ | LS$_4$ | |
| 0° | SYS | 0.229 | 0.684 | 2.775 | 4.076 | 0.597 | SYS | 0.235 | 0.612 | 1.857 | 2.510 | 0.564 |
| | COL | 0.605 | 1.439 | 2.773 | 4.068 | 0.582 | COL | 0.558 | 1.105 | 1.852 | 2.505 | 0.557 |
| | ABP | 1.535 | 5.940 | – | – | 0.929 | ABP | 1.105 | 2.989 | – | – | 0.781 |
| | ABA | 0.617 | 1.994 | – | – | 0.926 | ABA | 0.562 | 1.334 | – | – | 0.791 |
| | ABT | 0.228 | 0.712 | – | – | 0.605 | ABT | 0.233 | 0.644 | – | – | 0.570 |
| 15° | SYS | 0.176 | 0.561 | 2.433 | 3.521 | 0.591 | SYS | 0.188 | 0.522 | 1.652 | 2.215 | 0.579 |
| | COL | 0.556 | 1.286 | 2.426 | 3.516 | 0.574 | COL | 0.515 | 1.000 | 1.654 | 2.219 | 0.562 |
| | ABP | 1.858 | 7.062 | – | – | 0.931 | ABP | 1.254 | 3.336 | – | – | 0.763 |
| | ABA | 0.519 | 1.517 | – | – | 0.856 | ABA | 0.494 | 1.132 | – | – | 0.739 |
| | ABT | 0.174 | 0.580 | – | – | 0.606 | ABT | 0.190 | 0.540 | – | – | 0.600 |
| 30° | SYS | 0.167 | 0.559 | 2.265 | 3.260 | 0.590 | SYS | 0.194 | 0.529 | 1.555 | 2.063 | 0.576 |
| | COL | 0.535 | 1.213 | 2.253 | 3.236 | 0.565 | COL | 0.498 | 0.952 | 1.554 | 2.069 | 0.557 |
| | ABP | 2.491 | 9.505 | – | – | 0.928 | ABP | 1.636 | 4.476 | – | – | 0.747 |
| | ABA | 0.435 | 1.146 | – | – | 0.781 | ABA | 0.436 | 0.967 | – | – | 0.685 |
| | ABT | 0.167 | 0.582 | – | – | 0.618 | ABT | 0.193 | 0.547 | – | – | 0.598 |
| 45° | SYS | 0.185 | 0.636 | 2.245 | 3.225 | 0.590 | SYS | 0.220 | 0.578 | 1.528 | 2.026 | 0.557 |
| | COL | 0.536 | 1.211 | 2.244 | 3.218 | 0.564 | COL | 0.500 | 0.946 | 1.533 | 2.033 | 0.548 |
| | ABP | 3.069 | 11.867 | – | – | 0.895 | ABP | 2.095 | 6.072 | – | – | 0.734 |
| | ABA | 0.438 | 1.105 | – | – | 0.731 | ABA | 0.430 | 0.923 | – | – | 0.635 |
| | ABT | 0.184 | 0.684 | – | – | 0.634 | ABT | 0.222 | 0.615 | – | – | 0.571 |

[1] The dispersion value for the system is the average of the dispersions for LS$_1$, LS$_2$, LS$_3$, and LS$_4$.

**Table 4.** Fragilities for Seat Abutment Bridge Class for Different Skew Angles

| Skew angle | | Under FF ground motions | | | | | Under NF ground motions | | | | |
| --- | --- | --- | --- | --- | --- | --- | --- | --- | --- | --- | --- |
| | | $\lambda$ | | | | $\zeta^1$ | | $\lambda$ | | | $\zeta$ |
| | | $LS_1$ | $LS_2$ | $LS_3$ | $LS_4$ | | | $LS_1$ | $LS_2$ | $LS_3$ | $LS_4$ | |
| 0° | SYS | 0.133 | 0.578 | 2.540 | 3.828 | 0.639 | SYS | 0.126 | 0.540 | 1.752 | 2.405 | 0.647 |
| | COL | 0.518 | 1.286 | 2.559 | 3.826 | 0.572 | COL | 0.497 | 1.019 | 1.754 | 2.409 | 0.554 |
| | ABP | 1.583 | 4.467 | – | – | 0.768 | ABP | 1.231 | 2.851 | – | – | 0.700 |
| | ABA | 0.853 | 2.706 | – | – | 0.852 | ABA | 0.670 | 1.398 | – | – | 0.639 |
| | ABT | 0.221 | 0.660 | – | – | 0.645 | ABT | 0.215 | 0.637 | – | – | 0.642 |
| | UST | – | – | 14.527 | 29.154 | 0.997 | UST | – | – | 75.862 | 208.234 | 1.480 |
| | BRG | 0.141 | 0.944 | – | – | 0.848 | BRG | 0.140 | 0.958 | – | – | 1.034 |
| 15° | SYS | 0.122 | 0.509 | 2.227 | 3.303 | 0.638 | SYS | 0.099 | 0.479 | 1.578 | 2.159 | 0.676 |
| | COL | 0.482 | 1.159 | 2.251 | 3.319 | 0.570 | COL | 0.468 | 0.937 | 1.584 | 2.153 | 0.560 |
| | ABP | 1.895 | 5.976 | – | – | 0.793 | ABP | 1.400 | 3.515 | – | – | 0.708 |
| | ABA | 0.693 | 2.131 | – | – | 0.810 | ABA | 0.593 | 1.229 | – | – | 0.605 |
| | ABT | 0.199 | 0.567 | – | – | 0.638 | ABT | 0.196 | 0.556 | – | – | 0.644 |
| | UST | – | – | 9.462 | 18.223 | 0.947 | UST | – | – | 26.380 | 62.320 | 1.294 |
| | BRG | 0.130 | 0.865 | 0.000 | 0.000 | 0.872 | BRG | 0.108 | 0.969 | – | – | 1.152 |
| 30° | SYS | 0.122 | 0.520 | 2.088 | 3.096 | 0.637 | SYS | 0.098 | 0.488 | 1.510 | 2.065 | 0.681 |
| | COL | 0.466 | 1.105 | 2.124 | 3.114 | 0.564 | COL | 0.450 | 0.899 | 1.517 | 2.060 | 0.572 |
| | ABP | 2.671 | 9.798 | – | – | 0.838 | ABP | 1.968 | 5.842 | – | – | 0.746 |
| | ABA | 0.559 | 1.593 | – | – | 0.754 | ABA | 0.523 | 1.109 | – | – | 0.602 |
| | ABT | 0.202 | 0.581 | – | – | 0.633 | ABT | 0.202 | 0.562 | – | – | 0.639 |
| | UST | – | – | 6.665 | 12.424 | 0.910 | UST | – | – | 9.680 | 19.581 | 1.122 |
| | BRG | 0.128 | 0.866 | – | – | 0.858 | BRG | 0.102 | 0.939 | – | – | 1.092 |
| 45° | SYS | 0.124 | 0.573 | 1.999 | 2.964 | 0.628 | SYS | 0.117 | 0.525 | 1.461 | 2.010 | 0.651 |
| | COL | 0.457 | 1.078 | 2.065 | 3.019 | 0.566 | COL | 0.441 | 0.881 | 1.488 | 2.020 | 0.571 |
| | ABP | 3.488 | 14.271 | – | – | 0.857 | ABP | 2.778 | 9.772 | – | – | 0.778 |
| | ABA | 0.466 | 1.237 | – | – | 0.724 | ABA | 0.444 | 0.997 | – | – | 0.659 |
| | ABT | 0.227 | 0.689 | – | – | 0.630 | ABT | 0.241 | 0.642 | – | – | 0.612 |
| | UST | – | – | 4.873 | 8.739 | 0.852 | UST | – | – | 5.668 | 10.572 | 1.001 |
| | BRG | 0.129 | 0.875 | – | – | 0.813 | BRG | 0.121 | 0.859 | – | – | 0.930 |

[1]The dispersion value for the system is the average of the dispersions for $LS_1$, $LS_2$, $LS_3$, and $LS_4$.

**Table 5.** Skew Adjustment Factors for Seat Abutment Bridge Class under FF Motions

| Fragility | | LS$_1$ | LS$_2$ | LS$_3$ | LS$_4$ |
|---|---|---|---|---|---|
| SYS | $\lambda$ | $0.13–8\times10^{-4}\alpha–1\times10^{-5}\alpha^2$ | $0.58–6.1\times10^{-3}\alpha+1\times10^{-4}\alpha^2$ | $2.54–0.02\alpha+2\times10^{-4}\alpha^2$ | $3.82–0.04\alpha+4\times10^{-4}\alpha^2$ |
| | $\zeta$ | 0.753 | 0.632 | 0.581 | 0.576 |
| COL | $\lambda$ | $0.52–2.7\times10^{-3}\alpha+3\times10^{-5}\alpha^2$ | $1.28–9.5\times10^{-3}\alpha+1\times10^{-4}\alpha^2$ | $2.55–0.02\alpha+3\times10^{-4}\alpha^2$ | $3.82–0.04\alpha+5\times10^{-4}\alpha^2$ |
| | $\zeta$ | 0.568 | 0.568 | 0.568 | 0.568 |
| ABP | $\lambda$ | $1.56+0.02\alpha+6\times10^{-4}\alpha^2$ | $4.38+0.07\alpha+3\times10^{-3}\alpha^2$ | – | – |
| | $\zeta$ | 0.814 | 0.814 | – | – |
| ABA | $\lambda$ | $0.84–8.6\times10^{-3}\alpha$ | $2.71–0.04\alpha+2\times10^{-4}\alpha^2$ | – | – |
| | $\zeta$ | 0.785 | 0.785 | – | – |
| ABT | $\lambda$ | $0.22–2.2\times10^{-3}\alpha+5\times10^{-5}\alpha^2$ | $0.66–9.4\times10^{-3}\alpha+2\times10^{-4}\alpha^2$ | – | – |
| | $\zeta$ | 0.637 | 0.637 | – | – |
| UST | $\lambda$ | – | – | $14.46–0.38\alpha+3.6\times10^{-3}\alpha^2$ | $29.0–0.81\alpha+8.1\times10^{-3}\alpha^2$ |
| | $\zeta$ | – | – | 0.927 | 0.927 |
| BRG | $\lambda$ | $0.14–9\times10^{-3}\alpha+1\times10^{-5}\alpha^2$ | $0.94–5.7\times10^{-3}\alpha+1\times10^{-4}\alpha^2$ | – | – |
| | $\zeta$ | 0.848 | 0.848 | – | – |

**Table 6.** Skew Adjustment Factors for Seat Abutment Bridge Class under NF Motions

| Fragility | | LS$_1$ | LS$_2$ | LS$_3$ | LS$_4$ |
|---|---|---|---|---|---|
| SYS | $\lambda$ | $0.13–2.5\times10^{-3}\alpha+5\times10^{-5}\alpha^2$ | $0.54–5.1\times10^{-3}\alpha+1\times10^{-4}\alpha^2$ | $1.75–0.01\alpha+1\times10^{-4}\alpha^2$ | $2.4–0.02\alpha+2\times10^{-4}\alpha^2$ |
| | $\zeta$ | 0.859 | 0.658 | 0.571 | 0.567 |
| COL | $\lambda$ | $0.50–2.3\times10^{-3}\alpha+2\times10^{-5}\alpha^2$ | $1.02–6.2\times10^{-3}\alpha+7\times10^{-5}\alpha^2$ | $1.75–0.01\alpha+2\times10^{-4}\alpha^2$ | $2.4–0.02\alpha+2\times10^{-4}\alpha^2$ |
| | $\zeta$ | 0.564 | 0.564 | 0.564 | 0.564 |
| ABP | $\lambda$ | $1.22+2.7\times10^{-3}\alpha+7\times10^{-4}\alpha^2$ | $2.84–9.4\times10^{-3}\alpha+3.6\times10^{-3}\alpha^2$ | – | – |
| | $\zeta$ | 0.733 | 0.733 | – | – |
| ABA | $\lambda$ | $0.67–4.9\times10^{-3}\alpha$ | $1.4–0.01\alpha$ | – | – |
| | $\zeta$ | 0.626 | 0.626 | – | – |
| ABT | $\lambda$ | $0.22–2.4\times10^{-3}\alpha+7\times10^{-5}\alpha^2$ | $0.64–8\times10^{-3}\alpha+2\times10^{-4}\alpha^2$ | – | – |
| | $\zeta$ | 0.634 | 0.634 | – | – |
| UST | $\lambda$ | – | – | $74.85–3.79\alpha+0.05\alpha^2$ | $204.76–11.08\alpha+0.15\alpha^2$ |
| | $\zeta$ | – | – | 1.224 | 1.224 |
| BRG | $\lambda$ | $0.14–2.9\times10^{-3}\alpha+6\times10^{-5}\alpha^2$ | $0.96+2.4\times10^{-3}\alpha–1\times10^{-4}\alpha^2$ | – | – |
| | $\zeta$ | 1.052 | 1.052 | – | – |

**Table 7.** Skew Adjustment Factors for Diaphragm Abutment Bridge Class under FF Motions

| Fragility | | LS$_1$ | LS$_2$ | LS$_3$ | LS$_4$ |
|---|---|---|---|---|---|
| SYS | $\lambda$ | 0.23–4.5×10$^{-3}$$\alpha$+8×10$^{-5}$$\alpha^2$ | 0.68–0.01$\alpha$+2×10$^{-4}$$\alpha^2$ | 2.77–0.03$\alpha$+4×10$^{-4}$$\alpha^2$ | 4.07–0.04$\alpha$+6×10$^{-4}$$\alpha^2$ |
| | $\zeta$ | 0.621 | 0.600 | 0.574 | 0.574 |
| COL | $\lambda$ | 0.60–4.1×10$^{-3}$$\alpha$+6×10$^{-5}$$\alpha^2$ | 1.44–0.01$\alpha$+2×10$^{-4}$$\alpha^2$ | 2.77–0.03$\alpha$+4×10$^{-4}$$\alpha^2$ | 4.07–0.05$\alpha$+6×10$^{-4}$$\alpha^2$ |
| | $\zeta$ | 0.571 | 0.571 | 0.571 | 0.571 |
| ABP | $\lambda$ | 1.52+0.02$\alpha$+3×10$^{-4}$$\alpha^2$ | 5.87+0.07$\alpha$+1.4×10$^{-3}$$\alpha^2$ | – | – |
| | $\zeta$ | 0.921 | 0.921 | – | – |
| ABA | $\lambda$ | 0.62–9.2×10$^{-3}$$\alpha$+1×10$^{-4}$$\alpha^2$ | 2.01–0.04$\alpha$+5×10$^{-4}$$\alpha^2$ | – | – |
| | $\zeta$ | 0.824 | 0.824 | – | – |
| ABT | $\lambda$ | 0.22–4.5×10$^{-3}$$\alpha$+8×10$^{-5}$$\alpha^2$ | 0.71–0.01$\alpha$+3×10$^{-4}$$\alpha^2$ | – | – |
| | $\zeta$ | 0.616 | 0.616 | – | – |

**Table 8.** Skew Adjustment Factors for Diaphragm Abutment Bridge Class under NF Motions

| Fragility | | LS$_1$ | LS$_2$ | LS$_3$ | LS$_4$ |
|---|---|---|---|---|---|
| SYS | $\lambda$ | 0.23–3.9×10$^{-3}$$\alpha$+8×10$^{-5}$$\alpha^2$ | 0.61–7.6×10$^{-3}$$\alpha$+2×10$^{-4}$$\alpha^2$ | 1.86–0.02$\alpha$+2×10$^{-4}$$\alpha^2$ | 2.54–0.02$\alpha$+3×10$^{-4}$$\alpha^2$ |
| | $\zeta$ | 0.588 | 0.572 | 0.556 | 0.558 |
| COL | $\lambda$ | 0.56–3.5×10$^{-3}$$\alpha$+5×10$^{-5}$$\alpha^2$ | 1.1–8.4×10$^{-3}$$\alpha$+1×10$^{-4}$$\alpha^2$ | 1.85–0.02$\alpha$+2×10$^{-4}$$\alpha^2$ | 2.50–0.02$\alpha$+3×10$^{-4}$$\alpha^2$ |
| | $\zeta$ | 0.556 | 0.556 | 0.556 | 0.556 |
| ABP | $\lambda$ | 1.1+6.9×10$^{-3}$$\alpha$+3×10$^{-4}$$\alpha^2$ | 2.97+6.8×10$^{-3}$$\alpha$+1.4×10$^{-3}$$\alpha^2$ | – | – |
| | $\zeta$ | 756 | 756 | – | – |
| ABA | $\lambda$ | 0.56–6.1×10$^{-3}$$\alpha$+7×10$^{-5}$$\alpha^2$ | 1.34–0.02$\alpha$+2×10$^{-4}$$\alpha^2$ | – | – |
| | $\zeta$ | 0.713 | 0.713 | – | – |
| ABT | $\lambda$ | 0.23–3.9×10$^{-3}$$\alpha$+8×10$^{-5}$$\alpha^2$ | 0.64–9.1×10$^{-3}$$\alpha$+2×10$^{-4}$$\alpha^2$ | – | – |
| | $\zeta$ | 0.585 | 0.585 | – | – |

**List of Figures**

**Fig. 1.** Numerical Model of Skewed Bridges

**Fig. 2.** Comparison of Demolition Fragilities for Various Skew Angles

**Fig. 3.** Adjustment Factor for Seat Abutment Bridge Class at $LS_1$ under FF Ground Motions

**Fig. 4.** Comparison of Adjustment-Based and Simulation-Based System Fragility Curves under NF Ground Motions

**Fig. 5.** Comparison of Median Values of HAZUS and Adjusted System Fragility Curves under FF and NF Ground Motions.

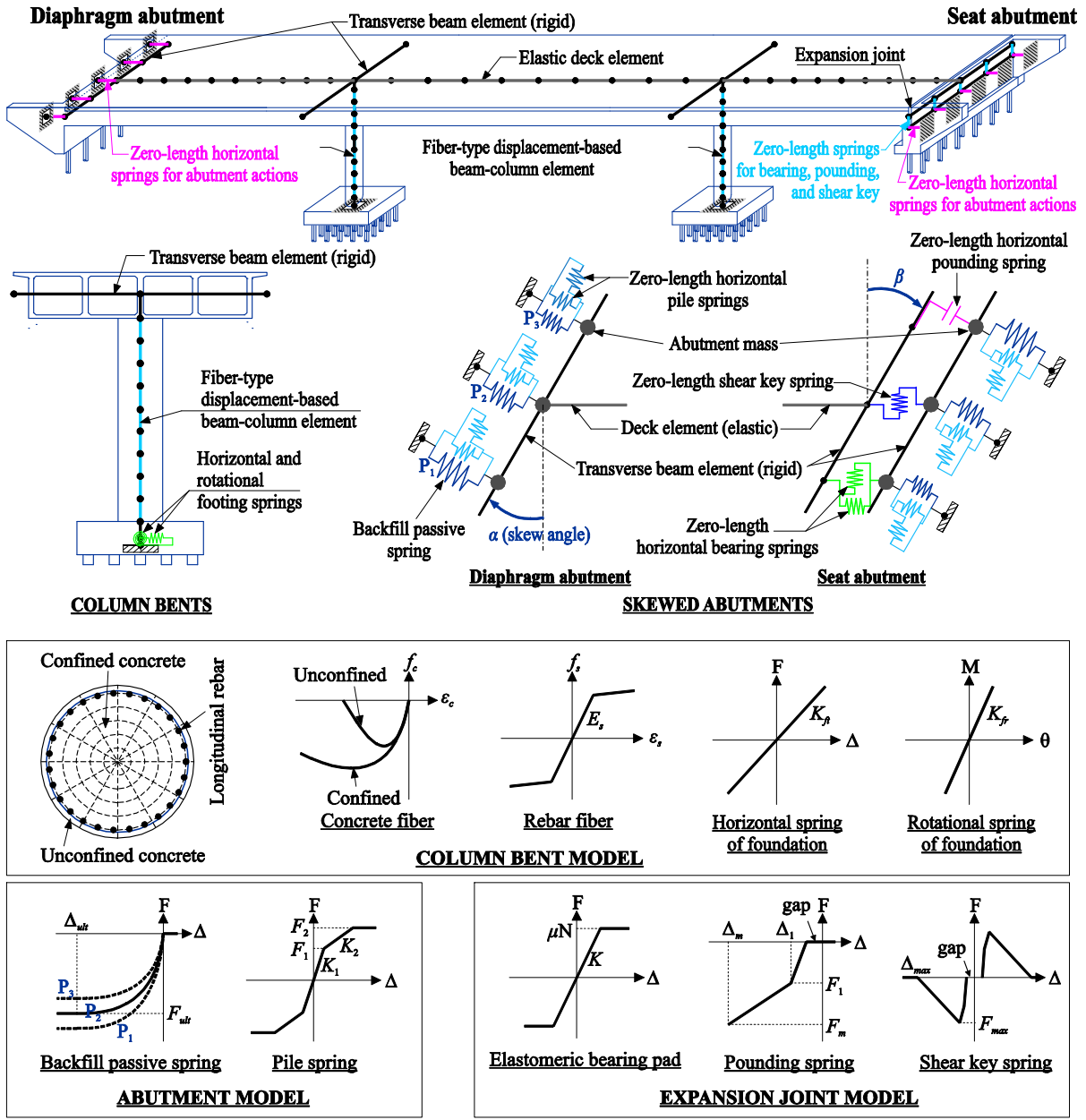

**Fig. 1.** Numerical Model of Skewed Bridges

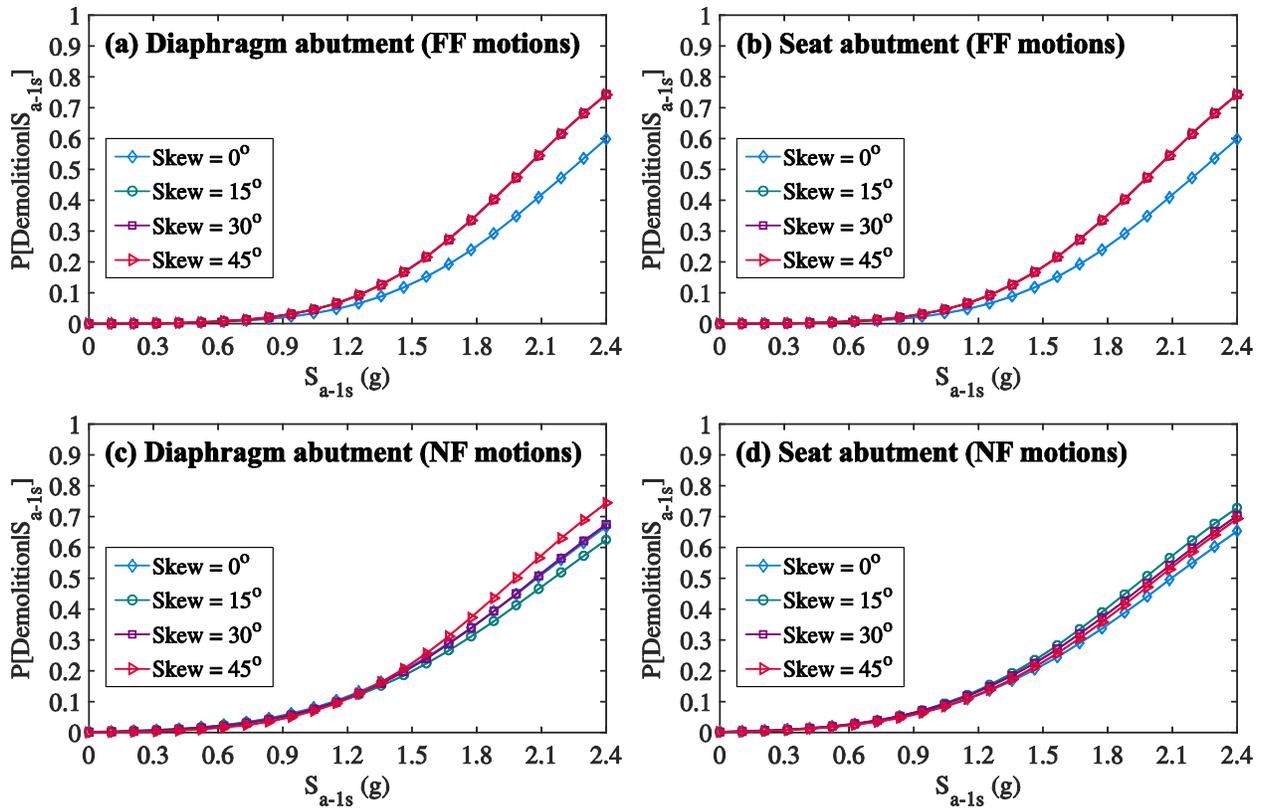

**Fig. 2.** Comparison of Demolition Fragilities for Various Skew Angles

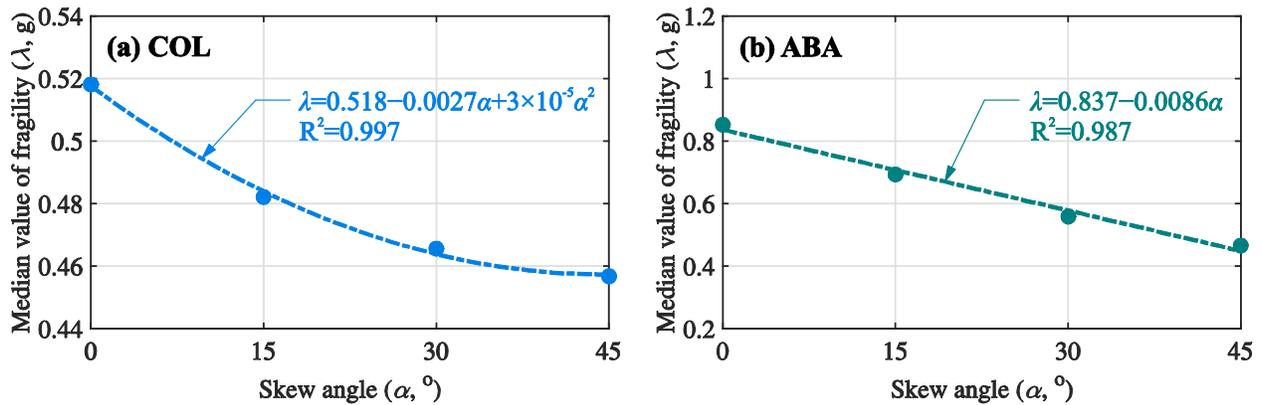

**Fig. 3.** Adjustment Factor for Seat Abutment Bridge Class at LS$_1$ under FF Ground Motions

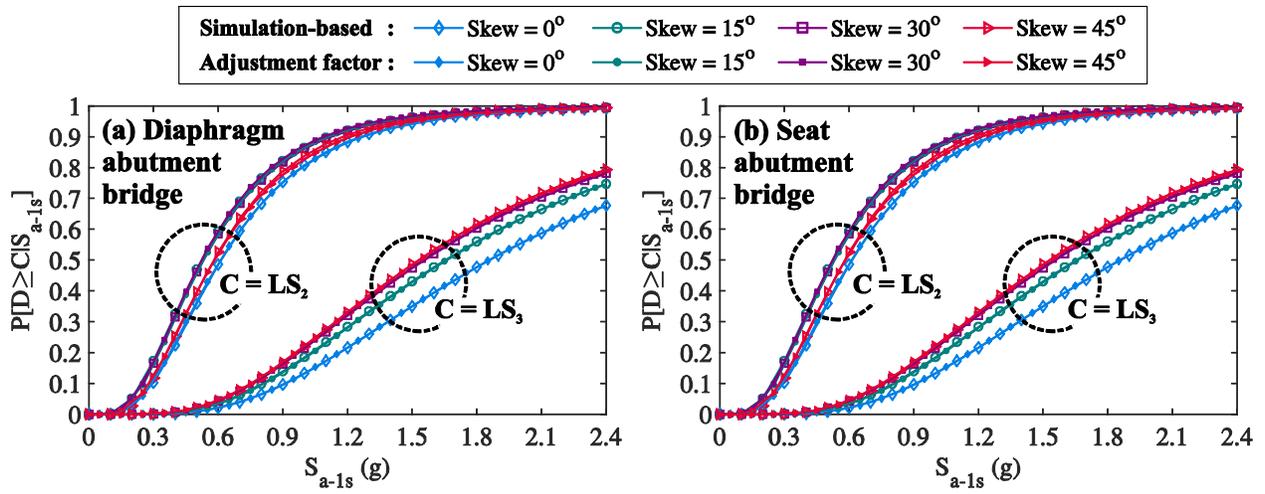

**Fig. 4.** Comparison of Adjustment-Based and Simulation-Based System Fragility Curves under NF Ground Motions

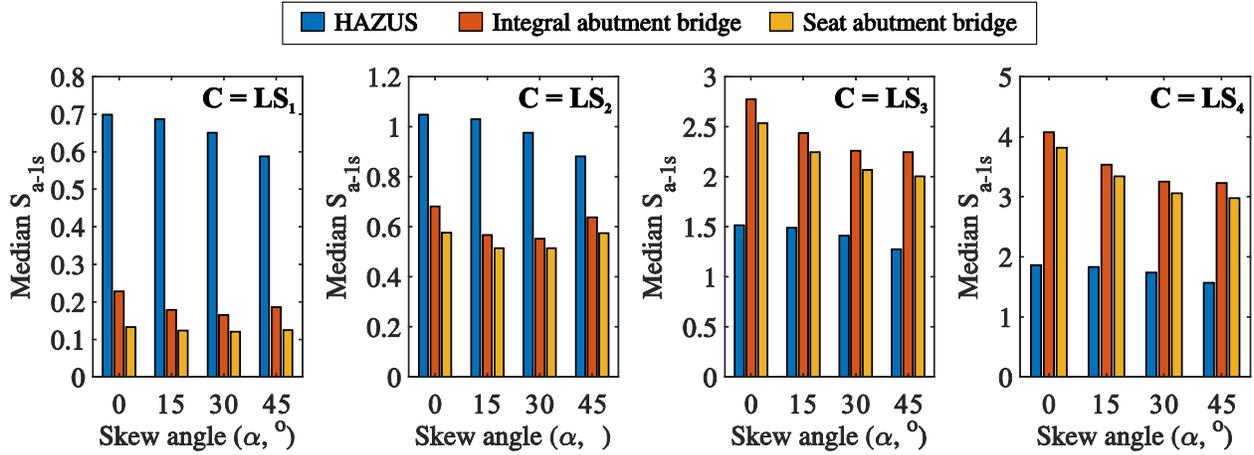

(a) Under FF motions

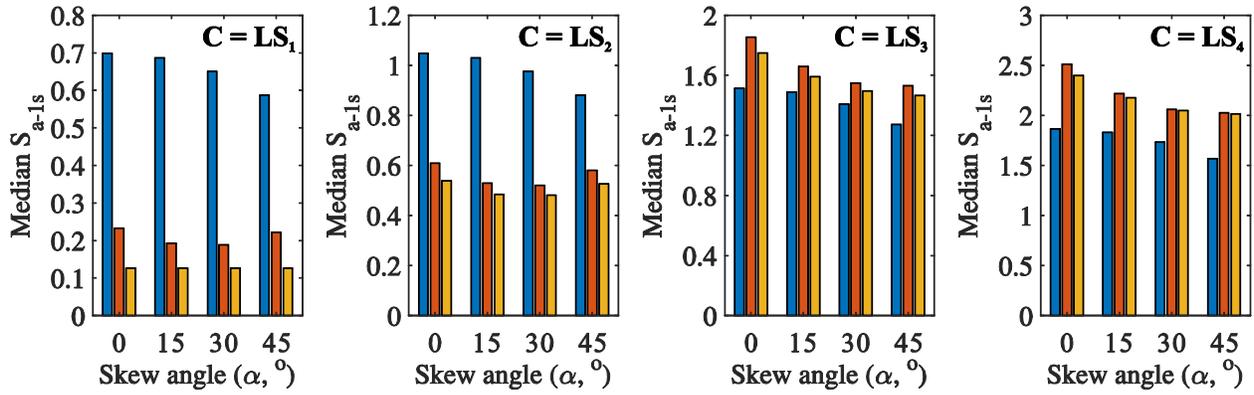

(b) Under NF motions

**Fig. 5.** Comparison of Median Values of HAZUS and Adjusted System Fragility Curves under FF and NF Ground Motions